\documentclass[aps,twocolumn,amsmath,amssymb,preprintnumbers,superscriptaddress,floatfix]{revtex4}

\usepackage[utf8]{inputenc}
\usepackage[T1]{fontenc}
\usepackage{newtxtext}
\usepackage[upint]{newtxmath}
\usepackage{microtype}
\usepackage{textcomp}
\usepackage{dsfont}
\usepackage{eucal}
\usepackage{siunitx}
\usepackage{soul}

\usepackage{enumerate}
\usepackage{amsfonts}
\usepackage{color}
\usepackage{soul}

\usepackage{graphicx}

\usepackage[colorlinks,allcolors=blue]{hyperref}
\usepackage[capitalize]{cleveref} 
\usepackage{cleveref}

\renewcommand{\vec}[1]{\boldsymbol{#1}}                           
\newcommand{\uv}[1]{\vec{e}_{#1}}       

\definecolor{DarkBlue}{rgb}{0,0,0.80}
\definecolor{DarkRed}{rgb}{0.80,0,0}
\definecolor{Purple}{rgb}{0.55,0,0.55}
\definecolor{Purple}{rgb}{0,0,0.8}

\DeclareMathOperator{\re}{Re}

\newcommand{\prlsection}[1]{\textbf{#1:}}

\let\epsilon\varepsilon

\begin{document}
\title{Superconducting proximity effect and long-ranged triplets in dirty metallic antiferromagnets}

\author{Eirik Holm Fyhn}
\affiliation{Center for Quantum Spintronics, Department of Physics, Norwegian \\ University of Science and Technology, NO-7491 Trondheim, Norway}
\author{Arne Brataas}
\affiliation{Center for Quantum Spintronics, Department of Physics, Norwegian \\ University of Science and Technology, NO-7491 Trondheim, Norway}
\author{Alireza Qaiumzadeh}
\affiliation{Center for Quantum Spintronics, Department of Physics, Norwegian \\ University of Science and Technology, NO-7491 Trondheim, Norway}
\author{Jacob Linder}
\affiliation{Center for Quantum Spintronics, Department of Physics, Norwegian \\ University of Science and Technology, NO-7491 Trondheim, Norway}

\date{\today}
\begin{abstract}
  \noindent Antiferromagnets have no net spin-splitting on the scale of the superconducting coherence length. Despite this, antiferromagnets have been observed to suppress superconductivity in a similar way as ferromagnets, a phenomenon that still lacks a clear understanding.
  We find that this effect can be explained by the role of impurities in antiferromagnets.
  Using quasiclassical Green's functions, we study the proximity effect and critical temperature in diffusive superconductor-metallic antiferromagnet bilayers.
  The non-magnetic impurities acquire an effective magnetic component in the antiferromagnet.
  This not only reduces the critical temperature but also separates the superconducting correlations into short-ranged and long-ranged components, similar to ferromagnetic proximity systems.
\end{abstract}
\maketitle
\prlsection{Introduction}
\label{sec:introduction}
Antiferromagnets and superconductors both have prominent roles in condensed matter physics~\cite{baltz2018,linder2015,fariborz2021,wadley2016,vaidya2020,wu2021,ma2021,chou2021,pan2022}.
Separately, they are both theoretically interesting due to their different types of quantum order~\cite{lee2006,haldane1983,sala2021,linder2019}.
They are also technologically useful: superconductors in part because of their perfect diamagnetism and dissipationless current~\cite{fossheim2004,jin2021}, and antiferromagnets because of their ultrafast dynamics~\cite{pimenov2009,baierl2016}, negligible stray-field and considerable magnetotransport effects~\cite{baltz2018}.
However, while materials with superconducting or magnetic properties can be interesting on their own, new physics and applications can be found in systems that combine both.
For instance, combining superconductivity and ferromagnetism in mesoscopic heterostructures is now a well-established method to produce odd-frequency superconductivity and long-range spin-triplet superconductivity~\cite{linder2019,bergeret2005}.
The latter can carry dissipationless spin-currents, giving superconductors a unique role in the field of spintronics~\cite{linder2015}.

Superconductor-antiferromagnet (SC-AF) heterostructures have been studied both theoretically and experimentally~\cite{bener2002,andersen_prb_05,zaitsev2010,moor_prb_12,kamra_prl_18,johnsen_prb_20,jakobsen2020,luntama2021,zhou2019,wu2013,hubener2002,bell2003}, but much less than their ferromagnetic counterparts.
As a result, much remains to be fully understood about SC-AF heterostructures.
For instance, experiments show that proximity to antiferromagnets can severely suppress the superconducting critical temperature~\cite{wu2013,hubener2002,bell2003}.
This suppression is much stronger than the prediction by the theoretical models which considered the AFs to be similar to normal metals due to their lack of uncompensated magnetic moments~\cite{hubener2002,bell2003,werthammer1963}.
In fact, the suppression has been reported to be even larger than the suppression seen in ferromagnetic junctions~\cite{wu2013}.
Various proposals have been suggested to explain this suppression, such as finite spin-splitting coming from uncompensated interfaces~\cite{bell2003}, the possibility of magnetic impurities having been infused into the superconductor during sample preparation~\cite{wu2013}, or the complex spin structure of the specific antiferromagnetic materials used in the experiments~\cite{hubener2002}.

More recently, in a theoretical study of superconductor-antiferromagnetic insulator bilayers with compensated interfaces, \citet{bobkov2022} suggested that a band-gap opening mechanism together with the induction of spin-triplet Cooper pairs could explain the suppression.
As these effects are smaller when the mean free path is shorter, they argued that the suppression would be larger for cleaner systems, but noted that a fully detailed analysis of the roles of impurities and AF length should consider a metallic AF.

Here, we study the proximity effect in diffusive superconductor~(SC)-antiferromagnetic metal~(AFM) bilayers using our newly derived quasiclassical framework~\cite{prb_submission}.
Interestingly, our results show that the suppression of superconductivity is not larger for clean systems, but that impurity scattering is in fact the dominant mechanism for superconductivity suppression in metallic AFs.
The reason is that the sublattice-spin coupling in the antiferromagnet gives the non-magnetic impurities an effective magnetic component.
These effective magnetic impurities are detrimental to superconductivity, except for spin-triplet superconductivity with spin aligned orthogonal to the Néel vector.
As a result, dirty AFMs work as superconductivity filters letting only spin-triplet superconductivity with orthogonal spin-projection to the Néel vector pass through.
After studying the critical temperature in SC/AFM bilayers, we show how the superconducting correlations penetrate into the antiferromagnetic metal, as well as the inverse proximity effect.
Moreover, we show how the long-range spin-triplet components can be induced by either uncompensated or magnetic interfaces with magnetic misalignment relative to the AFM Néel vector.

\prlsection{Theory}
To study SC-AFM bilayers, we employ the quasiclassical Keldysh formalism derived in~\cite{prb_submission}.
It is valid under the assumption that the Fermi wavelength is short compared to the coherence length and the mean free path, and the chemical potential, $\mu$, is much larger than all other energy scales in the system, except possibly the exchange energy between localized spins and conducting electrons, $J$.
Note that $\lvert J/\mu\rvert < 1$, since $\lvert \mu\rvert = \sqrt{J^2 + t^2}$, where $t$ is the hopping parameter evaluated at the Fermi surface.
We also assume the dirty limit, meaning that the system is diffusive, and that there is no electromagnetic vector potential.
In this case, the quasiclassical Green's function
$\check g$ solves~\cite{prb_submission}
\begin{align}
  i \nabla\cdot\check{\vec j} + \left[\tau_z(\varepsilon + i\delta) + \hat \Delta + \frac{iJ^2}{2\tau_\text{imp}\mu^2}\sigma_z\tau_z\check g\sigma_z\tau_z,\,\check g\right]
   = 0.
   \label{eq:usadel}
\end{align}
Here, $\check{\vec j}$ is the matrix current, $\tau_z$ and $\sigma_z$ are Pauli matrices in Nambu- and spin-space, respectively, $\varepsilon$ is energy, $\delta$ is the Dynes parameter, $\tau_\text{imp}$ is the elastic impurity scattering time and $\hat\Delta = \Delta i\tau_y$, under the assumption that the gap parameter $\Delta$ is real.
The spin-quantization axis is chosen to be parallel to the Néel vector, which is assumed homogeneous within the AFM.
We let the system be large in the directions parallel to the interface, such that the problem becomes an effective 1D problem.

The quasiclassical Green's function can be written
\begin{align}
 \check g = \begin{pmatrix}
   \hat g^R & \hat g^K \\ 0 & \hat g^A 
 \end{pmatrix},
\end{align}
where $\hat g^R$, $\hat g^A$ and $\hat g^K$ are the retarded, advanced and Keldysh Green's functions, respectively.
In thermal equilibrium, which is assumed here, it is sufficient to solve for $\hat g^R$, since $\hat g^A = -\tau_z(\hat g^R)^\dagger\tau_z$ and $\hat g^K = (\hat g^R - \hat g^A)\tanh(\beta\varepsilon/2)$.

\Cref{eq:usadel} is similar to the Usadel equation for normal metals~\cite{usadel1970} but is modified by the antiferromagnetic order in two important ways.
First, the expression for the matrix current is now
\begin{equation} 
  \check{\vec j} = 
    -D\check g\nabla\check g
    - \check g\left[\frac{J^2}{2\mu^2}\sigma_z\tau_z\check g\sigma_z\tau_z,\,\check{\vec j}\right],
  \label{eq:gen_curr}
\end{equation}
where $D$ is the diffusion constant.
The second way AFMs differ from normal metals is through the term proportional to $\sigma_z\tau_z\check g\sigma_z\tau_z$ in \cref{eq:usadel}.
This is exactly the way magnetic impurities enter the Usadel equation for normal metals~\cite{yokoyama2005}.
Hence, the antiferromagnetic order gives rise to effective magnetic impurities with scattering time equal to $\tau_\text{imp}\mu^2/J^2$.

The presence of $\tau_\text{imp}^{-1}$ in the equations requires some special care, as discussed in detail in Ref.~\cite{prb_submission}.
Since the impurity scattering rate, $\tau_\text{imp}^{-1}$, is small in the dirty limit, one should project onto only the long-ranged components of $\check g$ when $J^2/\mu^2 \approx 1$.
This is possible because some components become negligible in this limit of very strong exchange coupling.
Here, we consider smaller values of $J^2/\mu^2$.
Consequently, the effective magnetic scattering rate $J^2/(\mu^2\tau_\text{imp})$ is not necessarily large, and we must keep all components of $\check g$.

To model SC-AFM bilayers, we set $J = 0$ in the SC and $\Delta = 0$ in the AFM.
The dimensionless quantity $J/\mu$ is non-zero in the AFM, while the gap parameter in the SC is determined through the self-consistency equation~\cite{jacobsen2015},
\begin{align}
  \Delta = \frac{1}{4\operatorname{acosh}(\omega_D/\Delta_0)}\int_{0}^{\omega_D}\mathrm{d}{\varepsilon}\operatorname{Tr}\left[(\tau_x - i\tau_y)\hat g^K\right],
  \label{eq:gap}
\end{align}
where symmetries of the Green's function were used to write $\Delta$ as an integral over only positive energies, $\omega_D$ is a cutoff energy and $\Delta_0$ is a material-specific parameter defining the gap parameter in the bulk.
We set $\omega_D = 30\Delta_0$.

The two materials must be connected through a boundary condition, which is also derived in~\cite{prb_submission}.
We consider both compensated and uncompensated interfaces.
Let the interface be located at $x = 0$, and let $\check g_{SC} = \check g(0^-)$ and $\check g_{AF} = \check g(0^+)$ be the quasiclassical Green's functions on the superconducting and antiferromagnetic sides of the interface, respectively.
We similarly let $\check j_{SC}$ and $\check j_{AF}$ be the matrix current on the SC and AFM sides, respectively.
The general boundary condition for the matrix current going out of material $\alpha \in \{SC,AF\}$ and into material $\beta \in \{SC,AF\}$ is~\cite{prb_submission}
\begin{equation}
  \uv n \cdot \check{\vec j}_\alpha = \left[\hat{T}_{\alpha\beta} \check g_\beta\hat{T}_{\beta\alpha} + i\hat R_\alpha,\, \check g_\alpha\right].
  \label{eq:bc}
\end{equation}
where $\uv n$ is the outward-pointing normal vector, $\hat T_{\alpha\beta}$ is the tunneling matrix and $\hat R_\alpha$ is the reflection matrix.
For the case of a compensated interface, we assume that tunneling and reflection are independent of spin and sublattice.
In this case $\hat{T}_{\alpha\beta} = \hat{T}_{\beta\alpha}^* = t$ and $\hat R_\alpha$ are scalars.

In the case of an uncompensated interface, we assume that tunneling can only occur between the SC and the A-sublattice in the AFM.
In this case, we find from Ref.~\cite{prb_submission} that the tunneling matrix becomes
\begin{multline}
  T_{SC,AF} = T_{AF,SC}^\dagger = \frac t 2 \Bigl[\sqrt{1+J/\mu} + \sqrt{1-J/\mu} \\
  + \left(\sqrt{1+J/\mu} - \sqrt{1-J/\mu}\right)\tau_z\vec m \cdot \vec\sigma\Bigr].
\end{multline}
Here we have allowed for a possible misalignment between the magnetization direction in the AFM and the magnetization direction at the interface through the unit vector $\vec m$.
When the system is uncompensated, there should in general also be spin-dependent reflection.
For simplicity, we set the reflection matrix equal at both sides of the interface and equal to $\hat R_{SC} = \hat R_{AF} = r\tau_z\vec m\cdot \vec\sigma$.
Note that instead of an uncompensated interface, one can instead use a thin ferromagnetic (F) layer.
For instance, one could consider an SC/F/I/AFM structure, where the insulator (I) is used to reduce the exchange bias effect.
In this case, $\vec m$ would be the magnetization direction of the ferromagnet.

We solve \cref{eq:usadel,eq:gen_curr,eq:gap,eq:bc} numerically using the Ricatti-parametrization~\cite{eschrig2000,konstandin2005} and a collocation method~\cite{prb_submission}, and determine the matrix current by fixed-point iterations of \cref{eq:gen_curr}.
For simplicity, we set the diffusion constant to be equal in both materials.
We denote by $L_{AF}$ and $L_{SC}$ the lengths of the AFM and SC, respectively, and $\xi=\sqrt{D/\Delta_0}$ is the diffusive coherence length.
To find the critical temperature, we use the algorithm described in Ref.~\cite{ali2016}.

\begin{figure}[htpb]
  \centering
  \includegraphics[width=1.0\linewidth]{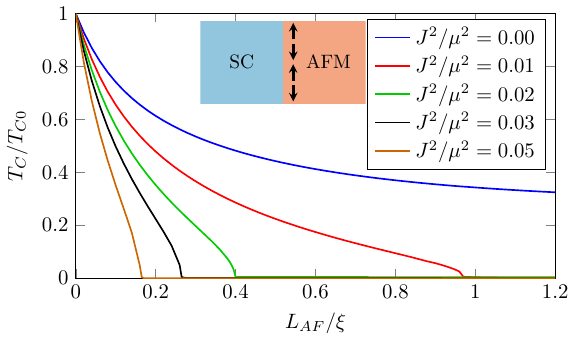}
  \caption{Critical temperature $T_C$ as a function of AFM length $L_{AF}$ normalized by the coherence length $\xi = \sqrt{D/\Delta_0}$. $T_{C0}$ is the bulk critical temperature. The inset shows a sketch of an SC/AFM bilayer with a compensated interface. The length of the superconductor is $L_{SC} = \xi$, the impurity scattering rate is $1/\tau_\text{imp} = 100\Delta_0$, the Dynes parameter is $\delta = 0.001\Delta_0$ and the tunneling amplitude is $t=3\sqrt{\Delta_0\xi}$.}%
  \label{fig:critTemp}
\end{figure}

\prlsection{Critical temperature}
We plot the critical temperature as a function of AFM length for various values of $J/\mu$ in \cref{fig:critTemp} for the case of a compensated interface.
As $J/\mu$ is increased, the critical temperature is reduced substantially, which is consistent with experiments~\cite{wu2013,hubener2002,bell2003}.
While $T_C$ decays slowly, reaching only around one-third of the bulk value in the SC/NM bilayers ($J/\mu = 0$), $T_C$ is reduced all the way to $0$ in the SC/AFM bilayers.
The AFM length needed to make $T_C$ vanish reduces with increasing $J/\mu$.
This can be understood from the effective magnetic impurities in the AFM, which has a scattering rate proportional to $J^2/\mu^2$.
It has long been known that even a small amount of magnetic impurities can strongly reduce the superconducting transition temperature~\cite{Abrikosov1960,hauser1966,jarell1988}.
The magnetic impurities give rise to spin-flip scatterings which break the spin-singlet Cooper pairs, thereby lowering the transition temperature.

Note that the maximal suppression of $T_C$ depends on the length of the superconductor and the magnitude of the tunneling amplitude.
When the superconductor is long compared to the coherence length, $T_C$ will be non-zero no matter how much the gap is suppressed near the interface.
Therefore, while one can observe total $T_C$ suppression for short superconductors, like in Ref.~\cite{hubener2002}, one should expect only a partial suppression, like in Refs.~\cite{wu2013,bell2003}, when the superconductor is long or the tunneling amplitude is small.
In the case of partial suppression, one should expect the minimal value to be reached when the length of the antiferromagnet reaches approximately the penetration depth of spin-singlet correlations, determined by $\tau_\text{imp}$ and $J/\mu$.

\begin{figure}[htpb]
  \centering
  \includegraphics[width=0.8\linewidth]{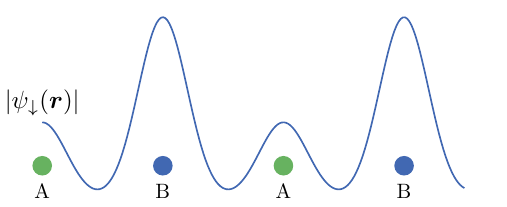}
  \caption{Exaggerated sketch of the spatial distribution of the conduction electron state with spin down. The overlap is larger with the B-sublattice than with the A-sublattice. As a result, the conduction band electrons with spin-down will be affected more strongly by non-magnetic impurities on the B-sublattice than by non-magnetic impurities on the A-sublattice.}%
  \label{fig:subLattSketch}
\end{figure}

To understand the origin of the effective magnetic impurities, consider the spatial distributions of the two degenerate spin-states of the antiferromagnetic conduction band.
The spin-down state is sketched in \cref{fig:subLattSketch}.
The spin-down (spin-up) state has larger amplitude on sublattice B (A) compared to sublattice.
As a result, non-magnetic impurities on sublattice B (A) act like superpositions of non-magnetic impurities and impurities with magnetization in the $-z$($+z$)-direction on the conduction band electrons. 
Therefore, electrons in the conduction band experience an effective magnetic impurity potential giving rise to spin-flip scattering described by the term proportional to $iJ^2/2\tau_\text{imp}\mu^2$ in \cref{eq:usadel}. 
The spin orientations of these impurities are locked along the direction of the Néel vector.
This gives rise to the possibility of long-ranged triplet correlations, as is shown in the following.

\prlsection{Proximity effect}
To study how the proximity effect is affected by the antiferromagnetic order, we consider the anomalous Green's function, $\operatorname{Tr}\left[(\tau_x - i\tau_y)\hat g^R\right] = f_0 + \vec f \cdot \vec\sigma$.
Here, $f_0$ describes the conventional spin-singlet superconducting correlations while $\vec f = (f_x, f_y, f_z)$ describes the spin-triplet correlations.
Note that since we are working with diffusive systems, the spin-singlet and spin-triplet correlations are also even and odd in frequency, respectively.

\begin{figure}[htpb]
  \centering
  \includegraphics[width=0.5\textwidth]{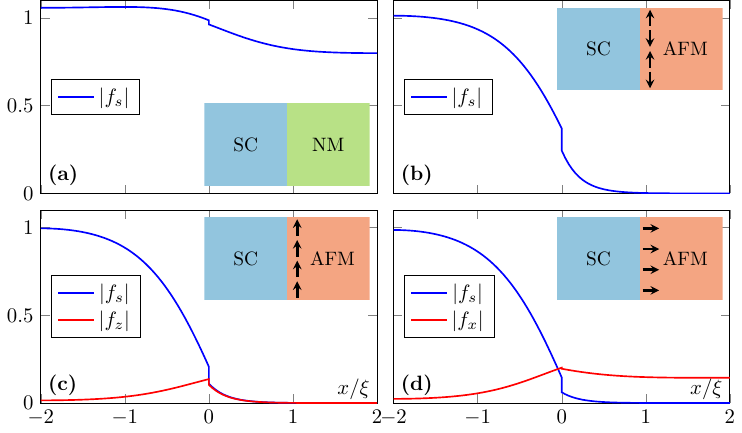}
  \caption{The non-zero components of the anomalous Green's function $f_0 + f_x\sigma_x+f_z\sigma_z$ evaluated at energy $\varepsilon = D/L_{AF}^2$ for various bilayer cases illustrated by the insets. The SC/AFM interface is at $x = 0$ and the Néel vector points in the $z$-direction. \textbf{(a)} SC/NM ($J/\mu = 0$), \textbf{(b)} SC/AFM with compensated interface, \textbf{(c)} SC/AFM with uncompensated interface aligned in the $z$-direction, \textbf{(d)} SC/AFM with uncompensated interface aligned in the $x$-direction. \textbf{(a)-(d)} have Dynes parameter $\delta = 0.001\Delta_0$, tunneling amplitude $t = 2\sqrt{\Delta_0\xi}$, temperature $T = 0.05T_{C0}$, SC length $L_{SC} = 2\xi$ and AFM length $L_{AF} = 2\xi$. \textbf{(b)-(d)} have $J^2/\mu^2 = 0.1$, $1/\tau_\text{imp} = 100\Delta_0$. \textbf{(c)-(d)} have $r =\Delta_0\xi$.}%
  \label{fig:anomGF}
\end{figure}

\Cref{fig:anomGF} shows the anomalous Green's function for various SC/AFM structures evaluated at $\varepsilon = D/L_{AF}^2$.
There is a large singlet component in the SC as expected.
In \cref{fig:anomGF}(a), the neighboring material is a normal metal, meaning that $J/\mu = 0$, and therefore the proximity induced $f_0$ penetrates deeply without significant decay.
On the other hand, in \cref{fig:anomGF}(b), the neighboring material has antiferromagnetic ordering with $J^2/\mu^2 = 0.1$ and $J^2/\mu^2\tau_\text{imp} = 10\Delta_0$.
In this case, the spin-singlet $f_0$ induced through the compensated interface decays over a much shorter length scale because of the effective magnetic impurities discussed above.
Additionally, $f_0$ is also more suppressed on the SC side, as expected from the $T_C$ results.

With an uncompensated interface, as shown in \cref{fig:anomGF}(c)-(d), spin-triplet correlations, $\vec f$, are also induced at the interface.
These correlations are aligned parallel to the magnetization direction of the interface.
When the correlations are parallel to the Néel-vector of the antiferromagnet, as in \cref{fig:anomGF}(c), they are affected by the magnetic impurities in the same way as the spin-singlet correlations, and therefore decay over the same length-scale.
However, when $\vec f$ is orthogonal to the Néel vector, as in \cref{fig:anomGF}(d), the spin-triplet correlations become long-ranged, decaying over a length scale that is the same as for a normal metal.

Thus, one can distinguish between short-ranged triplets and long-ranged triplets in diffusive AFMs, just as in FMs.
The reason why spin-singlet correlations and $\vec f \parallel \vec h$ are short-ranged in FMs with spin-splitting field $\vec h$ is because $\vec h$ induces an energy-difference between the electrons in the Cooper pairs, causing decoherence.
On the other hand, the spins of the two electrons in $\vec f \perp \vec h$ are both parallel to $\vec h$, such that they have the same wavelength as they propagate into the FM.
In diffusive antiferromagnets with Néel vector $\vec n$, the reason for the decoherence is non-magnetic impurities, but the effect is similar.
Spin-singlet $f_0$ and spin-triplet $\vec f \parallel \vec n$ are short-ranged while $\vec f \perp \vec n$ are long-ranged.

In order to compute the decay length associated with long-ranged and short-ranged correlations, we linearize the retarded component of \cref{eq:usadel} in the AFM.
We let $\hat g^R = \tau_z + i\tau_y f$.
To first order in $f$, we get from \cref{eq:gen_curr} that $\hat{\vec j}^R = -D\tau_x\nabla f/(1+J^2/\mu^2)$.
Inserting this into \cref{eq:usadel}, we get to first order in $f$ that
\begin{equation}
  \frac{D\nabla^2 f}{1+J^2/\mu^2} = -\left(2i\varepsilon - 2\delta - \frac{J^2}{\tau_\text{imp}\mu^2}\right)f + \frac{J^2}{\tau_\text{imp}\mu^2}\sigma_z f \sigma_z.
\end{equation}
Thus, if $\lim_{x\to\infty}f(x) = 0$ and $f(0) = a_0 + \vec a \cdot \vec \sigma$ for some constants $a_0$, and $\vec a = (a_x, a_y, a_z)$, then
  $
  f = (a_0 + a_z\sigma_z)\mathrm{e}^{ik_\parallel x} + (a_x\sigma_x + a_y\sigma_y)\mathrm{e}^{ik_\perp x},
  $
  where $k_\perp = \sqrt{2(1+J^2/\mu^2)(i\varepsilon - \delta)/D}$ and $k_\parallel = k_\perp\sqrt{(i\varepsilon - \delta - \tau_\text{imp}^{-1}J^2/\mu^2)/(i\varepsilon - \delta)}$,
such that the imaginary parts of $k_\parallel$ and $k_\perp$ are positive.

To find the decay lengths, we must take the imaginary parts of $k_\parallel$ and $k_\perp$.
When $\varepsilon \gg \delta$, the long-ranged correlations decay over a length scale equal to 
$
\lambda_\perp = 1/\operatorname{Im}(k_\perp)= \sqrt{D/[(1+J^2/\mu^2)\varepsilon]}
$.
This is on the same order as in a normal metal, $\lambda_\text{NM} = \sqrt{D/\varepsilon}$.
On the other hand, if $J^2/\mu^2\tau_\text{imp} \gg \varepsilon$, the short-ranged correlations decay over a length scale equal to
$
\lambda_\parallel = \mu l_\text{mfp}/[J\sqrt{6(1+J^2/\mu^2)}]
$,
which can be compared to the decay length for short-ranged correlations in ferromagnets~\cite{bergeret2005}, $\lambda_\text{FM} = \sqrt{D/h}$, where $h$ is the spin-splitting energy.
Here, $l_\text{mfp} = v_F\tau_\text{imp}$ is the mean free path, where $v_F$ is the Fermi velocity.

\begin{figure}[htpb]
  \centering
  \includegraphics[width=0.5\textwidth]{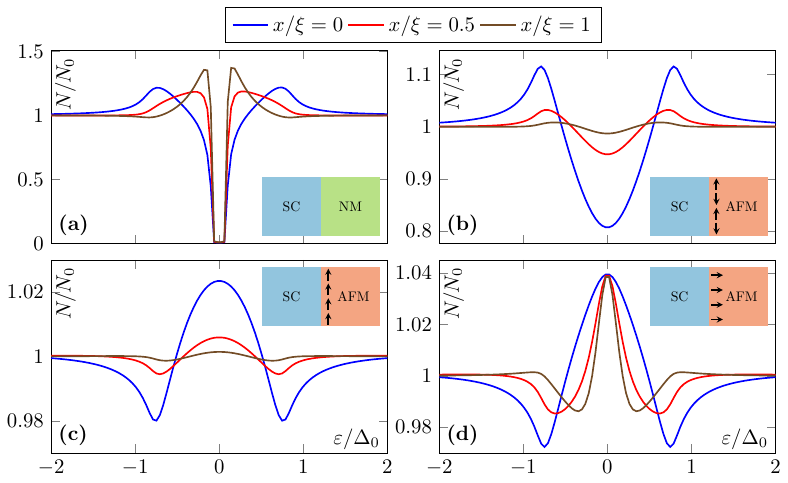}
  \caption{The local density of states, $N$, normalized by the normal state density of states $N_0$, at different positions inside the antiferromagnet.
  The parameters are the same as in \cref{fig:anomGF}, except that $J^2/\mu^2 = 0.01$ in \textbf{(b)}-\textbf{(d)}.
}
  \label{fig:dos}
\end{figure}

\prlsection{Local density of states}
In \cref{fig:dos} we show the normalized density of states, $N/N_0 = \re(\hat g_{11}^R + \hat g_{22}^R)/2$, as a function of energy at various positions with the same parameters as in \cref{fig:anomGF}, except that $J^2/\mu^2 = 0.01$ in \cref{fig:dos}(b)-(d).
The local density of states in the SC/NM bilayer is shown in \cref{fig:dos}(a), and one can see a minigap in the spectrum as expected~\cite{hammer2007}.
However, no minigap is present in the SC/AFM bilayers.
This is because the effective magnetic impurities act similarly to inelastic scattering for the spin-singlet correlations, leading to a much weaker suppression in the density of states, and a more smeared-out spectrum.
This is true also very close to the interface, as can be seen in \cref{fig:dos}(b).
As one moves away from the interface as $x=0$, the spectrum in the \cref{fig:dos}(b) rapidly becomes flatter.
This is in contrast to the spectrum in the SC/NM system, which retains the minigap also away from the interface.
The flatness of the spectrum away from the interface can be understood from the exponential decay coming from the effective magnetic impurities.

\Cref{fig:dos}(c) and \cref{fig:dos}(d) shows the local density of states for SC/AFM bilayers with short-ranged and long-ranged triplets respectively.
Both show a pronounced peak in the density of states at zero energy, similar to ferromagnetic systems with spin-triplet superconductivity~\cite{ali2017}.
Close to the interface, the spectrum in \cref{fig:dos}(c) is more smeared out than the spectrum in \cref{fig:dos}(d), probably because the effective magnetic impurities act on the short-ranged spin-triplet correlations in a similar way as they do on the spin-singlet correlations.
However, the biggest difference can be observed when going away from the interface.
While the spectrum in \cref{fig:dos}(c) becomes flatter, as expected from the exponential decay of the superconducting correlations, the spectrum in \cref{fig:dos}(d) transforms in the opposite way.
The zero-energy peak becomes sharper as the distance to the interface increases.

\prlsection{Conclusion}
We have theoretically studied diffusive SC/AFM bilayers with both compensated and uncompensated interfaces. 
We find a strong suppression of the critical temperature, consistent with experiments~\cite{wu2013,hubener2002,bell2003}.
This suppression can be explained in terms of effective magnetic impurities.
Non-magnetic impurities interact with conduction electrons in the AFM in a similar way as magnetic impurities in NMs.
Thus, we predict that cleaner AFMs will suppress superconductivity to a smaller degree, giving rise to higher critical temperatures.
The impurities in AFMs not only suppress spin-singlet superconductivity, and thereby $T_C$, but they also suppress spin-triplet correlations that are oriented parallel to the Néel vector.
As a result, spin-singlet correlations and spin-triplet correlations with parallel orientation are short-ranged, decaying exponentially over a length scale determined by the mean free path and the exchange energy between localized spins and conducting electrons.
In contrast, spin-triplet correlations with orientation orthogonal to the Néel vector are long-ranged.
They can penetrate as far as in normal metals.
Such long-range triplets can be induced by misaligned uncompensated interfaces, or by more complicated heterostructures.

\begin{acknowledgments}
This work was supported by the Research Council of Norway through grant 323766, and its Centres of Excellence funding scheme grant 262633 ``\emph{QuSpin}''. 
J.L. also acknowledges computational resources provided by Sigma2 - the National Infrastructure for High-Performance Computing and 
Data Storage in Norway from project no. NN9577K.
\end{acknowledgments}

\bibliography{bibliography}

\begin{thebibliography}{44}%
\makeatletter
\providecommand \@ifxundefined [1]{%
 \@ifx{#1\undefined}
}%
\providecommand \@ifnum [1]{%
 \ifnum #1\expandafter \@firstoftwo
 \else \expandafter \@secondoftwo
 \fi
}%
\providecommand \@ifx [1]{%
 \ifx #1\expandafter \@firstoftwo
 \else \expandafter \@secondoftwo
 \fi
}%
\providecommand \natexlab [1]{#1}%
\providecommand \enquote  [1]{``#1''}%
\providecommand \bibnamefont  [1]{#1}%
\providecommand \bibfnamefont [1]{#1}%
\providecommand \citenamefont [1]{#1}%
\providecommand \href@noop [0]{\@secondoftwo}%
\providecommand \href [0]{\begingroup \@sanitize@url \@href}%
\providecommand \@href[1]{\@@startlink{#1}\@@href}%
\providecommand \@@href[1]{\endgroup#1\@@endlink}%
\providecommand \@sanitize@url [0]{\catcode `\\12\catcode `\$12\catcode
  `\&12\catcode `\#12\catcode `\^12\catcode `\_12\catcode `\%12\relax}%
\providecommand \@@startlink[1]{}%
\providecommand \@@endlink[0]{}%
\providecommand \url  [0]{\begingroup\@sanitize@url \@url }%
\providecommand \@url [1]{\endgroup\@href {#1}{\urlprefix }}%
\providecommand \urlprefix  [0]{URL }%
\providecommand \Eprint [0]{\href }%
\providecommand \doibase [0]{http://dx.doi.org/}%
\providecommand \selectlanguage [0]{\@gobble}%
\providecommand \bibinfo  [0]{\@secondoftwo}%
\providecommand \bibfield  [0]{\@secondoftwo}%
\providecommand \translation [1]{[#1]}%
\providecommand \BibitemOpen [0]{}%
\providecommand \bibitemStop [0]{}%
\providecommand \bibitemNoStop [0]{.\EOS\space}%
\providecommand \EOS [0]{\spacefactor3000\relax}%
\providecommand \BibitemShut  [1]{\csname bibitem#1\endcsname}%
\let\auto@bib@innerbib\@empty
\bibitem [{\citenamefont {Baltz}\ \emph {et~al.}(2018)\citenamefont {Baltz},
  \citenamefont {Manchon}, \citenamefont {Tsoi}, \citenamefont {Moriyama},
  \citenamefont {Ono},\ and\ \citenamefont {Tserkovnyak}}]{baltz2018}%
  \BibitemOpen
  \bibfield  {author} {\bibinfo {author} {\bibfnamefont {V.}~\bibnamefont
  {Baltz}}, \bibinfo {author} {\bibfnamefont {A.}~\bibnamefont {Manchon}},
  \bibinfo {author} {\bibfnamefont {M.}~\bibnamefont {Tsoi}}, \bibinfo {author}
  {\bibfnamefont {T.}~\bibnamefont {Moriyama}}, \bibinfo {author}
  {\bibfnamefont {T.}~\bibnamefont {Ono}}, \ and\ \bibinfo {author}
  {\bibfnamefont {Y.}~\bibnamefont {Tserkovnyak}},\ }\href {\doibase
  10.1103/RevModPhys.90.015005} {\bibfield  {journal} {\bibinfo  {journal}
  {Rev. Mod. Phys.}\ }\textbf {\bibinfo {volume} {90}},\ \bibinfo {pages}
  {015005} (\bibinfo {year} {2018})}\BibitemShut {NoStop}%
\bibitem [{\citenamefont {Linder}\ and\ \citenamefont
  {Robinson}(2015)}]{linder2015}%
  \BibitemOpen
  \bibfield  {author} {\bibinfo {author} {\bibfnamefont {J.}~\bibnamefont
  {Linder}}\ and\ \bibinfo {author} {\bibfnamefont {J.~W.~A.}\ \bibnamefont
  {Robinson}},\ }\href {\doibase 10.1038/nphys3242} {\bibfield  {journal}
  {\bibinfo  {journal} {Nat. Phys.}\ }\textbf {\bibinfo {volume} {11}},\
  \bibinfo {pages} {307} (\bibinfo {year} {2015})}\BibitemShut {NoStop}%
\bibitem [{\citenamefont {Parhizgar}\ and\ \citenamefont
  {Black-Schaffer}(2021)}]{fariborz2021}%
  \BibitemOpen
  \bibfield  {author} {\bibinfo {author} {\bibfnamefont {F.}~\bibnamefont
  {Parhizgar}}\ and\ \bibinfo {author} {\bibfnamefont {A.~M.}\ \bibnamefont
  {Black-Schaffer}},\ }\href {\doibase 10.1103/PhysRevB.104.054507} {\bibfield
  {journal} {\bibinfo  {journal} {Phys. Rev. B}\ }\textbf {\bibinfo {volume}
  {104}},\ \bibinfo {pages} {054507} (\bibinfo {year} {2021})}\BibitemShut
  {NoStop}%
\bibitem [{\citenamefont {Wadley}\ \emph {et~al.}(2016)\citenamefont {Wadley},
  \citenamefont {Howells}, \citenamefont {Železný}, \citenamefont {Andrews},
  \citenamefont {Hills}, \citenamefont {Campion}, \citenamefont {Novák},
  \citenamefont {Olejník}, \citenamefont {Maccherozzi}, \citenamefont {Dhesi},
  \citenamefont {Martin}, \citenamefont {Wagner}, \citenamefont {Wunderlich},
  \citenamefont {Freimuth}, \citenamefont {Mokrousov}, \citenamefont {Kuneš},
  \citenamefont {Chauhan}, \citenamefont {Grzybowski}, \citenamefont
  {Rushforth}, \citenamefont {Edmonds}, \citenamefont {Gallagher},\ and\
  \citenamefont {Jungwirth}}]{wadley2016}%
  \BibitemOpen
  \bibfield  {author} {\bibinfo {author} {\bibfnamefont {P.}~\bibnamefont
  {Wadley}}, \bibinfo {author} {\bibfnamefont {B.}~\bibnamefont {Howells}},
  \bibinfo {author} {\bibfnamefont {J.}~\bibnamefont {Železný}}, \bibinfo
  {author} {\bibfnamefont {C.}~\bibnamefont {Andrews}}, \bibinfo {author}
  {\bibfnamefont {V.}~\bibnamefont {Hills}}, \bibinfo {author} {\bibfnamefont
  {R.~P.}\ \bibnamefont {Campion}}, \bibinfo {author} {\bibfnamefont
  {V.}~\bibnamefont {Novák}}, \bibinfo {author} {\bibfnamefont
  {K.}~\bibnamefont {Olejník}}, \bibinfo {author} {\bibfnamefont
  {F.}~\bibnamefont {Maccherozzi}}, \bibinfo {author} {\bibfnamefont {S.~S.}\
  \bibnamefont {Dhesi}}, \bibinfo {author} {\bibfnamefont {S.~Y.}\ \bibnamefont
  {Martin}}, \bibinfo {author} {\bibfnamefont {T.}~\bibnamefont {Wagner}},
  \bibinfo {author} {\bibfnamefont {J.}~\bibnamefont {Wunderlich}}, \bibinfo
  {author} {\bibfnamefont {F.}~\bibnamefont {Freimuth}}, \bibinfo {author}
  {\bibfnamefont {Y.}~\bibnamefont {Mokrousov}}, \bibinfo {author}
  {\bibfnamefont {J.}~\bibnamefont {Kuneš}}, \bibinfo {author} {\bibfnamefont
  {J.~S.}\ \bibnamefont {Chauhan}}, \bibinfo {author} {\bibfnamefont {M.~J.}\
  \bibnamefont {Grzybowski}}, \bibinfo {author} {\bibfnamefont {A.~W.}\
  \bibnamefont {Rushforth}}, \bibinfo {author} {\bibfnamefont {K.~W.}\
  \bibnamefont {Edmonds}}, \bibinfo {author} {\bibfnamefont {B.~L.}\
  \bibnamefont {Gallagher}}, \ and\ \bibinfo {author} {\bibfnamefont
  {T.}~\bibnamefont {Jungwirth}},\ }\href {\doibase 10.1126/science.aab1031}
  {\bibfield  {journal} {\bibinfo  {journal} {Science}\ }\textbf {\bibinfo
  {volume} {351}},\ \bibinfo {pages} {587} (\bibinfo {year}
  {2016})}\BibitemShut {NoStop}%
\bibitem [{\citenamefont {Vaidya}\ \emph {et~al.}(2020)\citenamefont {Vaidya},
  \citenamefont {Morley}, \citenamefont {van Tol}, \citenamefont {Liu},
  \citenamefont {Cheng}, \citenamefont {Brataas}, \citenamefont {Lederman},\
  and\ \citenamefont {del Barco}}]{vaidya2020}%
  \BibitemOpen
  \bibfield  {author} {\bibinfo {author} {\bibfnamefont {P.}~\bibnamefont
  {Vaidya}}, \bibinfo {author} {\bibfnamefont {S.~A.}\ \bibnamefont {Morley}},
  \bibinfo {author} {\bibfnamefont {J.}~\bibnamefont {van Tol}}, \bibinfo
  {author} {\bibfnamefont {Y.}~\bibnamefont {Liu}}, \bibinfo {author}
  {\bibfnamefont {R.}~\bibnamefont {Cheng}}, \bibinfo {author} {\bibfnamefont
  {A.}~\bibnamefont {Brataas}}, \bibinfo {author} {\bibfnamefont
  {D.}~\bibnamefont {Lederman}}, \ and\ \bibinfo {author} {\bibfnamefont
  {E.}~\bibnamefont {del Barco}},\ }\href {\doibase 10.1126/science.aaz4247}
  {\bibfield  {journal} {\bibinfo  {journal} {Science}\ }\textbf {\bibinfo
  {volume} {368}},\ \bibinfo {pages} {160} (\bibinfo {year}
  {2020})}\BibitemShut {NoStop}%
\bibitem [{\citenamefont {Wu}\ \emph {et~al.}(2021)\citenamefont {Wu},
  \citenamefont {Schwemmer}, \citenamefont {M\"uller}, \citenamefont
  {Consiglio}, \citenamefont {Sangiovanni}, \citenamefont {Di~Sante},
  \citenamefont {Iqbal}, \citenamefont {Hanke}, \citenamefont {Schnyder},
  \citenamefont {Denner}, \citenamefont {Fischer}, \citenamefont {Neupert},\
  and\ \citenamefont {Thomale}}]{wu2021}%
  \BibitemOpen
  \bibfield  {author} {\bibinfo {author} {\bibfnamefont {X.}~\bibnamefont
  {Wu}}, \bibinfo {author} {\bibfnamefont {T.}~\bibnamefont {Schwemmer}},
  \bibinfo {author} {\bibfnamefont {T.}~\bibnamefont {M\"uller}}, \bibinfo
  {author} {\bibfnamefont {A.}~\bibnamefont {Consiglio}}, \bibinfo {author}
  {\bibfnamefont {G.}~\bibnamefont {Sangiovanni}}, \bibinfo {author}
  {\bibfnamefont {D.}~\bibnamefont {Di~Sante}}, \bibinfo {author}
  {\bibfnamefont {Y.}~\bibnamefont {Iqbal}}, \bibinfo {author} {\bibfnamefont
  {W.}~\bibnamefont {Hanke}}, \bibinfo {author} {\bibfnamefont {A.~P.}\
  \bibnamefont {Schnyder}}, \bibinfo {author} {\bibfnamefont {M.~M.}\
  \bibnamefont {Denner}}, \bibinfo {author} {\bibfnamefont {M.~H.}\
  \bibnamefont {Fischer}}, \bibinfo {author} {\bibfnamefont {T.}~\bibnamefont
  {Neupert}}, \ and\ \bibinfo {author} {\bibfnamefont {R.}~\bibnamefont
  {Thomale}},\ }\href {\doibase 10.1103/PhysRevLett.127.177001} {\bibfield
  {journal} {\bibinfo  {journal} {Phys. Rev. Lett.}\ }\textbf {\bibinfo
  {volume} {127}},\ \bibinfo {pages} {177001} (\bibinfo {year}
  {2021})}\BibitemShut {NoStop}%
\bibitem [{\citenamefont {Ma}\ \emph {et~al.}(2022)\citenamefont {Ma},
  \citenamefont {Wang}, \citenamefont {Xie}, \citenamefont {Yang},
  \citenamefont {Wang}, \citenamefont {Zhou}, \citenamefont {Liu},
  \citenamefont {Yu}, \citenamefont {Zhao}, \citenamefont {Wang}, \citenamefont
  {Liu},\ and\ \citenamefont {Ma}}]{ma2021}%
  \BibitemOpen
  \bibfield  {author} {\bibinfo {author} {\bibfnamefont {L.}~\bibnamefont
  {Ma}}, \bibinfo {author} {\bibfnamefont {K.}~\bibnamefont {Wang}}, \bibinfo
  {author} {\bibfnamefont {Y.}~\bibnamefont {Xie}}, \bibinfo {author}
  {\bibfnamefont {X.}~\bibnamefont {Yang}}, \bibinfo {author} {\bibfnamefont
  {Y.}~\bibnamefont {Wang}}, \bibinfo {author} {\bibfnamefont {M.}~\bibnamefont
  {Zhou}}, \bibinfo {author} {\bibfnamefont {H.}~\bibnamefont {Liu}}, \bibinfo
  {author} {\bibfnamefont {X.}~\bibnamefont {Yu}}, \bibinfo {author}
  {\bibfnamefont {Y.}~\bibnamefont {Zhao}}, \bibinfo {author} {\bibfnamefont
  {H.}~\bibnamefont {Wang}}, \bibinfo {author} {\bibfnamefont {G.}~\bibnamefont
  {Liu}}, \ and\ \bibinfo {author} {\bibfnamefont {Y.}~\bibnamefont {Ma}},\
  }\href {\doibase 10.1103/PhysRevLett.128.167001} {\bibfield  {journal}
  {\bibinfo  {journal} {Phys. Rev. Lett.}\ }\textbf {\bibinfo {volume} {128}},\
  \bibinfo {pages} {167001} (\bibinfo {year} {2022})}\BibitemShut {NoStop}%
\bibitem [{\citenamefont {Chou}\ \emph {et~al.}(2021)\citenamefont {Chou},
  \citenamefont {Wu}, \citenamefont {Sau},\ and\ \citenamefont
  {Das~Sarma}}]{chou2021}%
  \BibitemOpen
  \bibfield  {author} {\bibinfo {author} {\bibfnamefont {Y.-Z.}\ \bibnamefont
  {Chou}}, \bibinfo {author} {\bibfnamefont {F.}~\bibnamefont {Wu}}, \bibinfo
  {author} {\bibfnamefont {J.~D.}\ \bibnamefont {Sau}}, \ and\ \bibinfo
  {author} {\bibfnamefont {S.}~\bibnamefont {Das~Sarma}},\ }\href {\doibase
  10.1103/PhysRevLett.127.187001} {\bibfield  {journal} {\bibinfo  {journal}
  {Phys. Rev. Lett.}\ }\textbf {\bibinfo {volume} {127}},\ \bibinfo {pages}
  {187001} (\bibinfo {year} {2021})}\BibitemShut {NoStop}%
\bibitem [{\citenamefont {Pan}\ \emph {et~al.}(2022)\citenamefont {Pan},
  \citenamefont {Le}, \citenamefont {He}, \citenamefont {Watzman},
  \citenamefont {Yao}, \citenamefont {Gooth}, \citenamefont {Heremans},
  \citenamefont {Sun},\ and\ \citenamefont {Felser}}]{pan2022}%
  \BibitemOpen
  \bibfield  {author} {\bibinfo {author} {\bibfnamefont {Y.}~\bibnamefont
  {Pan}}, \bibinfo {author} {\bibfnamefont {C.}~\bibnamefont {Le}}, \bibinfo
  {author} {\bibfnamefont {B.}~\bibnamefont {He}}, \bibinfo {author}
  {\bibfnamefont {S.~J.}\ \bibnamefont {Watzman}}, \bibinfo {author}
  {\bibfnamefont {M.}~\bibnamefont {Yao}}, \bibinfo {author} {\bibfnamefont
  {J.}~\bibnamefont {Gooth}}, \bibinfo {author} {\bibfnamefont {J.~P.}\
  \bibnamefont {Heremans}}, \bibinfo {author} {\bibfnamefont {Y.}~\bibnamefont
  {Sun}}, \ and\ \bibinfo {author} {\bibfnamefont {C.}~\bibnamefont {Felser}},\
  }\href {\doibase 10.1038/s41563-021-01149-2} {\bibfield  {journal} {\bibinfo
  {journal} {Nat. Mater.}\ }\textbf {\bibinfo {volume} {21}},\ \bibinfo {pages}
  {203} (\bibinfo {year} {2022})}\BibitemShut {NoStop}%
\bibitem [{\citenamefont {Lee}\ \emph {et~al.}(2006)\citenamefont {Lee},
  \citenamefont {Nagaosa},\ and\ \citenamefont {Wen}}]{lee2006}%
  \BibitemOpen
  \bibfield  {author} {\bibinfo {author} {\bibfnamefont {P.~A.}\ \bibnamefont
  {Lee}}, \bibinfo {author} {\bibfnamefont {N.}~\bibnamefont {Nagaosa}}, \ and\
  \bibinfo {author} {\bibfnamefont {X.-G.}\ \bibnamefont {Wen}},\ }\href
  {\doibase 10.1103/RevModPhys.78.17} {\bibfield  {journal} {\bibinfo
  {journal} {Rev. Mod. Phys.}\ }\textbf {\bibinfo {volume} {78}},\ \bibinfo
  {pages} {17} (\bibinfo {year} {2006})}\BibitemShut {NoStop}%
\bibitem [{\citenamefont {Haldane}(1983)}]{haldane1983}%
  \BibitemOpen
  \bibfield  {author} {\bibinfo {author} {\bibfnamefont {F.~D.~M.}\
  \bibnamefont {Haldane}},\ }\href {\doibase 10.1103/PhysRevLett.50.1153}
  {\bibfield  {journal} {\bibinfo  {journal} {Phys. Rev. Lett.}\ }\textbf
  {\bibinfo {volume} {50}},\ \bibinfo {pages} {1153} (\bibinfo {year}
  {1983})}\BibitemShut {NoStop}%
\bibitem [{\citenamefont {Sala}\ \emph {et~al.}(2021)\citenamefont {Sala},
  \citenamefont {Stone}, \citenamefont {Rai}, \citenamefont {May},
  \citenamefont {Laurell}, \citenamefont {Garlea}, \citenamefont {Butch},
  \citenamefont {Lumsden}, \citenamefont {Ehlers}, \citenamefont {Pokharel},
  \citenamefont {Podlesnyak}, \citenamefont {Mandrus}, \citenamefont {Parker},
  \citenamefont {Okamoto}, \citenamefont {Halász},\ and\ \citenamefont
  {Christianson}}]{sala2021}%
  \BibitemOpen
  \bibfield  {author} {\bibinfo {author} {\bibfnamefont {G.}~\bibnamefont
  {Sala}}, \bibinfo {author} {\bibfnamefont {M.~B.}\ \bibnamefont {Stone}},
  \bibinfo {author} {\bibfnamefont {B.~K.}\ \bibnamefont {Rai}}, \bibinfo
  {author} {\bibfnamefont {A.~F.}\ \bibnamefont {May}}, \bibinfo {author}
  {\bibfnamefont {P.}~\bibnamefont {Laurell}}, \bibinfo {author} {\bibfnamefont
  {V.~O.}\ \bibnamefont {Garlea}}, \bibinfo {author} {\bibfnamefont {N.~P.}\
  \bibnamefont {Butch}}, \bibinfo {author} {\bibfnamefont {M.~D.}\ \bibnamefont
  {Lumsden}}, \bibinfo {author} {\bibfnamefont {G.}~\bibnamefont {Ehlers}},
  \bibinfo {author} {\bibfnamefont {G.}~\bibnamefont {Pokharel}}, \bibinfo
  {author} {\bibfnamefont {A.}~\bibnamefont {Podlesnyak}}, \bibinfo {author}
  {\bibfnamefont {D.}~\bibnamefont {Mandrus}}, \bibinfo {author} {\bibfnamefont
  {D.~S.}\ \bibnamefont {Parker}}, \bibinfo {author} {\bibfnamefont
  {S.}~\bibnamefont {Okamoto}}, \bibinfo {author} {\bibfnamefont {G.~B.}\
  \bibnamefont {Halász}}, \ and\ \bibinfo {author} {\bibfnamefont {A.~D.}\
  \bibnamefont {Christianson}},\ }\href {\doibase 10.1038/s41467-020-20335-5}
  {\bibfield  {journal} {\bibinfo  {journal} {Nat. Commun.}\ }\textbf {\bibinfo
  {volume} {12}},\ \bibinfo {pages} {171} (\bibinfo {year} {2021})}\BibitemShut
  {NoStop}%
\bibitem [{\citenamefont {Linder}\ and\ \citenamefont
  {Balatsky}(2019)}]{linder2019}%
  \BibitemOpen
  \bibfield  {author} {\bibinfo {author} {\bibfnamefont {J.}~\bibnamefont
  {Linder}}\ and\ \bibinfo {author} {\bibfnamefont {A.~V.}\ \bibnamefont
  {Balatsky}},\ }\href {\doibase 10.1103/RevModPhys.91.045005} {\bibfield
  {journal} {\bibinfo  {journal} {Rev. Mod. Phys.}\ }\textbf {\bibinfo {volume}
  {91}},\ \bibinfo {pages} {045005} (\bibinfo {year} {2019})}\BibitemShut
  {NoStop}%
\bibitem [{\citenamefont {Fossheim}\ and\ \citenamefont
  {Sudb{\o}}(2004)}]{fossheim2004}%
  \BibitemOpen
  \bibfield  {author} {\bibinfo {author} {\bibfnamefont {K.}~\bibnamefont
  {Fossheim}}\ and\ \bibinfo {author} {\bibfnamefont {A.}~\bibnamefont
  {Sudb{\o}}},\ }\href@noop {} {\emph {\bibinfo {title} {Superconductivity:
  physics and applications}}}\ (\bibinfo  {publisher} {John Wiley \& Sons},\
  \bibinfo {year} {2004})\BibitemShut {NoStop}%
\bibitem [{\citenamefont {Jin}\ \emph {et~al.}(2021)\citenamefont {Jin},
  \citenamefont {Sheng}, \citenamefont {Bi}, \citenamefont {Song},
  \citenamefont {Liu}, \citenamefont {Chen}, \citenamefont {Li}, \citenamefont
  {Deng}, \citenamefont {Zhang}, \citenamefont {Zheng}, \citenamefont {Coombs},
  \citenamefont {Shen}, \citenamefont {Zhu}, \citenamefont {Zhao},
  \citenamefont {Wang}, \citenamefont {Xiang}, \citenamefont {Tang},
  \citenamefont {Ren}, \citenamefont {Xu}, \citenamefont {Shi}, \citenamefont
  {Islam}, \citenamefont {Guo},\ and\ \citenamefont {Zhu}}]{jin2021}%
  \BibitemOpen
  \bibfield  {author} {\bibinfo {author} {\bibfnamefont {J.}~\bibnamefont
  {Jin}}, \bibinfo {author} {\bibfnamefont {G.}~\bibnamefont {Sheng}}, \bibinfo
  {author} {\bibfnamefont {Y.}~\bibnamefont {Bi}}, \bibinfo {author}
  {\bibfnamefont {Y.}~\bibnamefont {Song}}, \bibinfo {author} {\bibfnamefont
  {X.}~\bibnamefont {Liu}}, \bibinfo {author} {\bibfnamefont {X.}~\bibnamefont
  {Chen}}, \bibinfo {author} {\bibfnamefont {Q.}~\bibnamefont {Li}}, \bibinfo
  {author} {\bibfnamefont {Z.}~\bibnamefont {Deng}}, \bibinfo {author}
  {\bibfnamefont {W.}~\bibnamefont {Zhang}}, \bibinfo {author} {\bibfnamefont
  {J.}~\bibnamefont {Zheng}}, \bibinfo {author} {\bibfnamefont
  {T.}~\bibnamefont {Coombs}}, \bibinfo {author} {\bibfnamefont
  {B.}~\bibnamefont {Shen}}, \bibinfo {author} {\bibfnamefont {J.}~\bibnamefont
  {Zhu}}, \bibinfo {author} {\bibfnamefont {Y.}~\bibnamefont {Zhao}}, \bibinfo
  {author} {\bibfnamefont {J.}~\bibnamefont {Wang}}, \bibinfo {author}
  {\bibfnamefont {B.}~\bibnamefont {Xiang}}, \bibinfo {author} {\bibfnamefont
  {Y.}~\bibnamefont {Tang}}, \bibinfo {author} {\bibfnamefont {L.}~\bibnamefont
  {Ren}}, \bibinfo {author} {\bibfnamefont {Y.}~\bibnamefont {Xu}}, \bibinfo
  {author} {\bibfnamefont {J.}~\bibnamefont {Shi}}, \bibinfo {author}
  {\bibfnamefont {M.~R.}\ \bibnamefont {Islam}}, \bibinfo {author}
  {\bibfnamefont {Y.}~\bibnamefont {Guo}}, \ and\ \bibinfo {author}
  {\bibfnamefont {J.}~\bibnamefont {Zhu}},\ }\href {\doibase
  10.1109/TASC.2021.3108740} {\bibfield  {journal} {\bibinfo  {journal} {IEEE
  Trans. Appl. Supercond.}\ }\textbf {\bibinfo {volume} {31}},\ \bibinfo
  {pages} {1} (\bibinfo {year} {2021})}\BibitemShut {NoStop}%
\bibitem [{\citenamefont {Pimenov}\ \emph {et~al.}(2009)\citenamefont
  {Pimenov}, \citenamefont {Shuvaev}, \citenamefont {Loidl}, \citenamefont
  {Schrettle}, \citenamefont {Mukhin}, \citenamefont {Travkin}, \citenamefont
  {Ivanov},\ and\ \citenamefont {Balbashov}}]{pimenov2009}%
  \BibitemOpen
  \bibfield  {author} {\bibinfo {author} {\bibfnamefont {A.}~\bibnamefont
  {Pimenov}}, \bibinfo {author} {\bibfnamefont {A.}~\bibnamefont {Shuvaev}},
  \bibinfo {author} {\bibfnamefont {A.}~\bibnamefont {Loidl}}, \bibinfo
  {author} {\bibfnamefont {F.}~\bibnamefont {Schrettle}}, \bibinfo {author}
  {\bibfnamefont {A.~A.}\ \bibnamefont {Mukhin}}, \bibinfo {author}
  {\bibfnamefont {V.~D.}\ \bibnamefont {Travkin}}, \bibinfo {author}
  {\bibfnamefont {V.~Y.}\ \bibnamefont {Ivanov}}, \ and\ \bibinfo {author}
  {\bibfnamefont {A.~M.}\ \bibnamefont {Balbashov}},\ }\href {\doibase
  10.1103/PhysRevLett.102.107203} {\bibfield  {journal} {\bibinfo  {journal}
  {Phys. Rev. Lett.}\ }\textbf {\bibinfo {volume} {102}},\ \bibinfo {pages}
  {107203} (\bibinfo {year} {2009})}\BibitemShut {NoStop}%
\bibitem [{\citenamefont {Baierl}\ \emph {et~al.}(2016)\citenamefont {Baierl},
  \citenamefont {Mentink}, \citenamefont {Hohenleutner}, \citenamefont {Braun},
  \citenamefont {Do}, \citenamefont {Lange}, \citenamefont {Sell},
  \citenamefont {Fiebig}, \citenamefont {Woltersdorf}, \citenamefont
  {Kampfrath},\ and\ \citenamefont {Huber}}]{baierl2016}%
  \BibitemOpen
  \bibfield  {author} {\bibinfo {author} {\bibfnamefont {S.}~\bibnamefont
  {Baierl}}, \bibinfo {author} {\bibfnamefont {J.~H.}\ \bibnamefont {Mentink}},
  \bibinfo {author} {\bibfnamefont {M.}~\bibnamefont {Hohenleutner}}, \bibinfo
  {author} {\bibfnamefont {L.}~\bibnamefont {Braun}}, \bibinfo {author}
  {\bibfnamefont {T.-M.}\ \bibnamefont {Do}}, \bibinfo {author} {\bibfnamefont
  {C.}~\bibnamefont {Lange}}, \bibinfo {author} {\bibfnamefont
  {A.}~\bibnamefont {Sell}}, \bibinfo {author} {\bibfnamefont {M.}~\bibnamefont
  {Fiebig}}, \bibinfo {author} {\bibfnamefont {G.}~\bibnamefont {Woltersdorf}},
  \bibinfo {author} {\bibfnamefont {T.}~\bibnamefont {Kampfrath}}, \ and\
  \bibinfo {author} {\bibfnamefont {R.}~\bibnamefont {Huber}},\ }\href
  {\doibase 10.1103/PhysRevLett.117.197201} {\bibfield  {journal} {\bibinfo
  {journal} {Phys. Rev. Lett.}\ }\textbf {\bibinfo {volume} {117}},\ \bibinfo
  {pages} {197201} (\bibinfo {year} {2016})}\BibitemShut {NoStop}%
\bibitem [{\citenamefont {Bergeret}\ \emph {et~al.}(2005)\citenamefont
  {Bergeret}, \citenamefont {Volkov},\ and\ \citenamefont
  {Efetov}}]{bergeret2005}%
  \BibitemOpen
  \bibfield  {author} {\bibinfo {author} {\bibfnamefont {F.~S.}\ \bibnamefont
  {Bergeret}}, \bibinfo {author} {\bibfnamefont {A.~F.}\ \bibnamefont
  {Volkov}}, \ and\ \bibinfo {author} {\bibfnamefont {K.~B.}\ \bibnamefont
  {Efetov}},\ }\href {\doibase 10.1103/RevModPhys.77.1321} {\bibfield
  {journal} {\bibinfo  {journal} {Rev. Mod. Phys.}\ }\textbf {\bibinfo {volume}
  {77}},\ \bibinfo {pages} {1321} (\bibinfo {year} {2005})}\BibitemShut
  {NoStop}%
\bibitem [{\citenamefont {bener}\ \emph {et~al.}(2002)\citenamefont {bener},
  \citenamefont {Tikhonov}, \citenamefont {Garifullin}, \citenamefont
  {Westerholt},\ and\ \citenamefont {Zabel}}]{bener2002}%
  \BibitemOpen
  \bibfield  {author} {\bibinfo {author} {\bibfnamefont {M.~H.}\ \bibnamefont
  {bener}}, \bibinfo {author} {\bibfnamefont {D.}~\bibnamefont {Tikhonov}},
  \bibinfo {author} {\bibfnamefont {I.~A.}\ \bibnamefont {Garifullin}},
  \bibinfo {author} {\bibfnamefont {K.}~\bibnamefont {Westerholt}}, \ and\
  \bibinfo {author} {\bibfnamefont {H.}~\bibnamefont {Zabel}},\ }\href
  {\doibase 10.1088/0953-8984/14/37/305} {\bibfield  {journal} {\bibinfo
  {journal} {Journal of Physics: Condensed Matter}\ }\textbf {\bibinfo {volume}
  {14}},\ \bibinfo {pages} {8687} (\bibinfo {year} {2002})}\BibitemShut
  {NoStop}%
\bibitem [{\citenamefont {Andersen}\ \emph {et~al.}(2005)\citenamefont
  {Andersen}, \citenamefont {Bobkova}, \citenamefont {Hirschfeld},\ and\
  \citenamefont {Barash}}]{andersen_prb_05}%
  \BibitemOpen
  \bibfield  {author} {\bibinfo {author} {\bibfnamefont {B.~M.}\ \bibnamefont
  {Andersen}}, \bibinfo {author} {\bibfnamefont {I.~V.}\ \bibnamefont
  {Bobkova}}, \bibinfo {author} {\bibfnamefont {P.~J.}\ \bibnamefont
  {Hirschfeld}}, \ and\ \bibinfo {author} {\bibfnamefont {Y.~S.}\ \bibnamefont
  {Barash}},\ }\href {\doibase 10.1103/PhysRevB.72.184510} {\bibfield
  {journal} {\bibinfo  {journal} {Phys. Rev. B}\ }\textbf {\bibinfo {volume}
  {72}},\ \bibinfo {pages} {184510} (\bibinfo {year} {2005})}\BibitemShut
  {NoStop}%
\bibitem [{\citenamefont {Zaitsev}\ \emph {et~al.}(2010)\citenamefont
  {Zaitsev}, \citenamefont {Ovsyannikov}, \citenamefont {Constantinian},
  \citenamefont {Kislinskiĭ}, \citenamefont {Shadrin}, \citenamefont
  {Borisenko},\ and\ \citenamefont {Komissinskiy}}]{zaitsev2010}%
  \BibitemOpen
  \bibfield  {author} {\bibinfo {author} {\bibfnamefont {A.}~\bibnamefont
  {Zaitsev}}, \bibinfo {author} {\bibfnamefont {G.~A.}\ \bibnamefont
  {Ovsyannikov}}, \bibinfo {author} {\bibfnamefont {K.~Y.}\ \bibnamefont
  {Constantinian}}, \bibinfo {author} {\bibfnamefont {Y.~V.}\ \bibnamefont
  {Kislinskiĭ}}, \bibinfo {author} {\bibfnamefont {A.~V.}\ \bibnamefont
  {Shadrin}}, \bibinfo {author} {\bibfnamefont {I.~V.}\ \bibnamefont
  {Borisenko}}, \ and\ \bibinfo {author} {\bibfnamefont {P.~V.}\ \bibnamefont
  {Komissinskiy}},\ }\href {\doibase 10.1134/S1063776110020172} {\bibfield
  {journal} {\bibinfo  {journal} {J. Exp. Theor. Phys.}\ }\textbf {\bibinfo
  {volume} {110}},\ \bibinfo {pages} {336} (\bibinfo {year}
  {2010})}\BibitemShut {NoStop}%
\bibitem [{\citenamefont {Moor}\ \emph {et~al.}(2012)\citenamefont {Moor},
  \citenamefont {Volkov},\ and\ \citenamefont {Efetov}}]{moor_prb_12}%
  \BibitemOpen
  \bibfield  {author} {\bibinfo {author} {\bibfnamefont {A.}~\bibnamefont
  {Moor}}, \bibinfo {author} {\bibfnamefont {A.~F.}\ \bibnamefont {Volkov}}, \
  and\ \bibinfo {author} {\bibfnamefont {K.~B.}\ \bibnamefont {Efetov}},\
  }\href {\doibase 10.1103/PhysRevB.85.014523} {\bibfield  {journal} {\bibinfo
  {journal} {Phys. Rev. B}\ }\textbf {\bibinfo {volume} {85}},\ \bibinfo
  {pages} {014523} (\bibinfo {year} {2012})}\BibitemShut {NoStop}%
\bibitem [{\citenamefont {Kamra}\ \emph {et~al.}(2018)\citenamefont {Kamra},
  \citenamefont {Rezaei},\ and\ \citenamefont {Belzig}}]{kamra_prl_18}%
  \BibitemOpen
  \bibfield  {author} {\bibinfo {author} {\bibfnamefont {A.}~\bibnamefont
  {Kamra}}, \bibinfo {author} {\bibfnamefont {A.}~\bibnamefont {Rezaei}}, \
  and\ \bibinfo {author} {\bibfnamefont {W.}~\bibnamefont {Belzig}},\ }\href
  {\doibase 10.1103/PhysRevLett.121.247702} {\bibfield  {journal} {\bibinfo
  {journal} {Phys. Rev. Lett.}\ }\textbf {\bibinfo {volume} {121}},\ \bibinfo
  {pages} {247702} (\bibinfo {year} {2018})}\BibitemShut {NoStop}%
\bibitem [{\citenamefont {Johnsen}\ \emph {et~al.}(2021)\citenamefont
  {Johnsen}, \citenamefont {Jacobsen},\ and\ \citenamefont
  {Linder}}]{johnsen_prb_20}%
  \BibitemOpen
  \bibfield  {author} {\bibinfo {author} {\bibfnamefont {L.~G.}\ \bibnamefont
  {Johnsen}}, \bibinfo {author} {\bibfnamefont {S.~H.}\ \bibnamefont
  {Jacobsen}}, \ and\ \bibinfo {author} {\bibfnamefont {J.}~\bibnamefont
  {Linder}},\ }\href {\doibase 10.1103/PhysRevB.103.L060505} {\bibfield
  {journal} {\bibinfo  {journal} {Phys. Rev. B}\ }\textbf {\bibinfo {volume}
  {103}},\ \bibinfo {pages} {L060505} (\bibinfo {year} {2021})}\BibitemShut
  {NoStop}%
\bibitem [{\citenamefont {Jakobsen}\ \emph {et~al.}(2020)\citenamefont
  {Jakobsen}, \citenamefont {Naess}, \citenamefont {Dutta}, \citenamefont
  {Brataas},\ and\ \citenamefont {Qaiumzadeh}}]{jakobsen2020}%
  \BibitemOpen
  \bibfield  {author} {\bibinfo {author} {\bibfnamefont {M.~F.}\ \bibnamefont
  {Jakobsen}}, \bibinfo {author} {\bibfnamefont {K.~B.}\ \bibnamefont {Naess}},
  \bibinfo {author} {\bibfnamefont {P.}~\bibnamefont {Dutta}}, \bibinfo
  {author} {\bibfnamefont {A.}~\bibnamefont {Brataas}}, \ and\ \bibinfo
  {author} {\bibfnamefont {A.}~\bibnamefont {Qaiumzadeh}},\ }\href {\doibase
  10.1103/PhysRevB.102.140504} {\bibfield  {journal} {\bibinfo  {journal}
  {Phys. Rev. B}\ }\textbf {\bibinfo {volume} {102}},\ \bibinfo {pages}
  {140504(R)} (\bibinfo {year} {2020})}\BibitemShut {NoStop}%
\bibitem [{\citenamefont {Luntama}\ \emph {et~al.}(2021)\citenamefont
  {Luntama}, \citenamefont {T\"orm\"a},\ and\ \citenamefont
  {Lado}}]{luntama2021}%
  \BibitemOpen
  \bibfield  {author} {\bibinfo {author} {\bibfnamefont {S.~S.}\ \bibnamefont
  {Luntama}}, \bibinfo {author} {\bibfnamefont {P.}~\bibnamefont {T\"orm\"a}},
  \ and\ \bibinfo {author} {\bibfnamefont {J.~L.}\ \bibnamefont {Lado}},\
  }\href {\doibase 10.1103/PhysRevResearch.3.L012021} {\bibfield  {journal}
  {\bibinfo  {journal} {Phys. Rev. Research}\ }\textbf {\bibinfo {volume}
  {3}},\ \bibinfo {pages} {L012021} (\bibinfo {year} {2021})}\BibitemShut
  {NoStop}%
\bibitem [{\citenamefont {Zhou}\ \emph {et~al.}(2019)\citenamefont {Zhou},
  \citenamefont {Lan}, \citenamefont {Ye}, \citenamefont {Feng}, \citenamefont
  {Zhai}, \citenamefont {Gong}, \citenamefont {Wang}, \citenamefont {Zhao},\
  and\ \citenamefont {Xu}}]{zhou2019}%
  \BibitemOpen
  \bibfield  {author} {\bibinfo {author} {\bibfnamefont {X.}~\bibnamefont
  {Zhou}}, \bibinfo {author} {\bibfnamefont {M.}~\bibnamefont {Lan}}, \bibinfo
  {author} {\bibfnamefont {Y.}~\bibnamefont {Ye}}, \bibinfo {author}
  {\bibfnamefont {Y.}~\bibnamefont {Feng}}, \bibinfo {author} {\bibfnamefont
  {X.}~\bibnamefont {Zhai}}, \bibinfo {author} {\bibfnamefont {L.}~\bibnamefont
  {Gong}}, \bibinfo {author} {\bibfnamefont {H.}~\bibnamefont {Wang}}, \bibinfo
  {author} {\bibfnamefont {J.}~\bibnamefont {Zhao}}, \ and\ \bibinfo {author}
  {\bibfnamefont {Y.}~\bibnamefont {Xu}},\ }\href {\doibase
  10.1209/0295-5075/125/37001} {\bibfield  {journal} {\bibinfo  {journal}
  {EPL}\ }\textbf {\bibinfo {volume} {125}},\ \bibinfo {pages} {37001}
  (\bibinfo {year} {2019})}\BibitemShut {NoStop}%
\bibitem [{\citenamefont {Wu}\ \emph {et~al.}(2013)\citenamefont {Wu},
  \citenamefont {Yang}, \citenamefont {Guo}, \citenamefont {Wu},\ and\
  \citenamefont {Qiu}}]{wu2013}%
  \BibitemOpen
  \bibfield  {author} {\bibinfo {author} {\bibfnamefont {B.~L.}\ \bibnamefont
  {Wu}}, \bibinfo {author} {\bibfnamefont {Y.~M.}\ \bibnamefont {Yang}},
  \bibinfo {author} {\bibfnamefont {Z.~B.}\ \bibnamefont {Guo}}, \bibinfo
  {author} {\bibfnamefont {Y.~H.}\ \bibnamefont {Wu}}, \ and\ \bibinfo {author}
  {\bibfnamefont {J.~J.}\ \bibnamefont {Qiu}},\ }\href {\doibase
  10.1063/1.4824891} {\bibfield  {journal} {\bibinfo  {journal} {Appl. Phys.
  Lett.}\ }\textbf {\bibinfo {volume} {103}},\ \bibinfo {pages} {152602}
  (\bibinfo {year} {2013})}\BibitemShut {NoStop}%
\bibitem [{\citenamefont {Hübener}\ \emph {et~al.}(2002)\citenamefont
  {Hübener}, \citenamefont {Tikhonov}, \citenamefont {Garifullin},
  \citenamefont {Westerholt},\ and\ \citenamefont {Zabel}}]{hubener2002}%
  \BibitemOpen
  \bibfield  {author} {\bibinfo {author} {\bibfnamefont {M.}~\bibnamefont
  {Hübener}}, \bibinfo {author} {\bibfnamefont {D.}~\bibnamefont {Tikhonov}},
  \bibinfo {author} {\bibfnamefont {I.~A.}\ \bibnamefont {Garifullin}},
  \bibinfo {author} {\bibfnamefont {K.}~\bibnamefont {Westerholt}}, \ and\
  \bibinfo {author} {\bibfnamefont {H.}~\bibnamefont {Zabel}},\ }\href
  {\doibase 10.1088/0953-8984/14/37/305} {\bibfield  {journal} {\bibinfo
  {journal} {J. Phys.: Condens. Matter}\ }\textbf {\bibinfo {volume} {14}},\
  \bibinfo {pages} {8687} (\bibinfo {year} {2002})}\BibitemShut {NoStop}%
\bibitem [{\citenamefont {Bell}\ \emph {et~al.}(2003)\citenamefont {Bell},
  \citenamefont {Tarte}, \citenamefont {Burnell}, \citenamefont {Leung},
  \citenamefont {Kang},\ and\ \citenamefont {Blamire}}]{bell2003}%
  \BibitemOpen
  \bibfield  {author} {\bibinfo {author} {\bibfnamefont {C.}~\bibnamefont
  {Bell}}, \bibinfo {author} {\bibfnamefont {E.~J.}\ \bibnamefont {Tarte}},
  \bibinfo {author} {\bibfnamefont {G.}~\bibnamefont {Burnell}}, \bibinfo
  {author} {\bibfnamefont {C.~W.}\ \bibnamefont {Leung}}, \bibinfo {author}
  {\bibfnamefont {D.-J.}\ \bibnamefont {Kang}}, \ and\ \bibinfo {author}
  {\bibfnamefont {M.~G.}\ \bibnamefont {Blamire}},\ }\href {\doibase
  10.1103/PhysRevB.68.144517} {\bibfield  {journal} {\bibinfo  {journal} {Phys.
  Rev. B}\ }\textbf {\bibinfo {volume} {68}},\ \bibinfo {pages} {144517}
  (\bibinfo {year} {2003})}\BibitemShut {NoStop}%
\bibitem [{\citenamefont {Werthamer}(1963)}]{werthammer1963}%
  \BibitemOpen
  \bibfield  {author} {\bibinfo {author} {\bibfnamefont {N.~R.}\ \bibnamefont
  {Werthamer}},\ }\href {\doibase 10.1103/PhysRev.132.2440} {\bibfield
  {journal} {\bibinfo  {journal} {Phys. Rev.}\ }\textbf {\bibinfo {volume}
  {132}},\ \bibinfo {pages} {2440} (\bibinfo {year} {1963})}\BibitemShut
  {NoStop}%
\bibitem [{\citenamefont {Bobkov}\ \emph {et~al.}(2022)\citenamefont {Bobkov},
  \citenamefont {Bobkova}, \citenamefont {Bobkov},\ and\ \citenamefont
  {Kamra}}]{bobkov2022}%
  \BibitemOpen
  \bibfield  {author} {\bibinfo {author} {\bibfnamefont {G.~A.}\ \bibnamefont
  {Bobkov}}, \bibinfo {author} {\bibfnamefont {I.~V.}\ \bibnamefont {Bobkova}},
  \bibinfo {author} {\bibfnamefont {A.~M.}\ \bibnamefont {Bobkov}}, \ and\
  \bibinfo {author} {\bibfnamefont {A.}~\bibnamefont {Kamra}},\ }\href
  {\doibase 10.1103/PhysRevB.106.144512} {\bibfield  {journal} {\bibinfo
  {journal} {Phys. Rev. B}\ }\textbf {\bibinfo {volume} {106}},\ \bibinfo
  {pages} {144512} (\bibinfo {year} {2022})}\BibitemShut {NoStop}%
\bibitem [{\citenamefont {Fyhn}\ \emph {et~al.}(2023)\citenamefont {Fyhn},
  \citenamefont {Brataas}, \citenamefont {Qaiumzadeh},\ and\ \citenamefont
  {Linder}}]{prb_submission}%
  \BibitemOpen
  \bibfield  {author} {\bibinfo {author} {\bibfnamefont {E.~H.}\ \bibnamefont
  {Fyhn}}, \bibinfo {author} {\bibfnamefont {A.}~\bibnamefont {Brataas}},
  \bibinfo {author} {\bibfnamefont {A.}~\bibnamefont {Qaiumzadeh}}, \ and\
  \bibinfo {author} {\bibfnamefont {J.}~\bibnamefont {Linder}},\ }\href
  {\doibase 10.1103/PhysRevB.107.174503} {\bibfield  {journal} {\bibinfo
  {journal} {Phys. Rev. B}\ }\textbf {\bibinfo {volume} {107}},\ \bibinfo
  {pages} {174503} (\bibinfo {year} {2023})}\BibitemShut {NoStop}%
\bibitem [{\citenamefont {Usadel}(1970)}]{usadel1970}%
  \BibitemOpen
  \bibfield  {author} {\bibinfo {author} {\bibfnamefont {K.~D.}\ \bibnamefont
  {Usadel}},\ }\href {\doibase 10.1103/PhysRevLett.25.507} {\bibfield
  {journal} {\bibinfo  {journal} {Phys. Rev. Lett.}\ }\textbf {\bibinfo
  {volume} {25}},\ \bibinfo {pages} {507} (\bibinfo {year} {1970})}\BibitemShut
  {NoStop}%
\bibitem [{\citenamefont {Yokoyama}\ \emph {et~al.}(2005)\citenamefont
  {Yokoyama}, \citenamefont {Tanaka}, \citenamefont {Golubov}, \citenamefont
  {Inoue},\ and\ \citenamefont {Asano}}]{yokoyama2005}%
  \BibitemOpen
  \bibfield  {author} {\bibinfo {author} {\bibfnamefont {T.}~\bibnamefont
  {Yokoyama}}, \bibinfo {author} {\bibfnamefont {Y.}~\bibnamefont {Tanaka}},
  \bibinfo {author} {\bibfnamefont {A.~A.}\ \bibnamefont {Golubov}}, \bibinfo
  {author} {\bibfnamefont {J.}~\bibnamefont {Inoue}}, \ and\ \bibinfo {author}
  {\bibfnamefont {Y.}~\bibnamefont {Asano}},\ }\href {\doibase
  10.1103/PhysRevB.71.094506} {\bibfield  {journal} {\bibinfo  {journal} {Phys.
  Rev. B}\ }\textbf {\bibinfo {volume} {71}},\ \bibinfo {pages} {094506}
  (\bibinfo {year} {2005})}\BibitemShut {NoStop}%
\bibitem [{\citenamefont {Jacobsen}\ \emph {et~al.}(2015)\citenamefont
  {Jacobsen}, \citenamefont {Ouassou},\ and\ \citenamefont
  {Linder}}]{jacobsen2015}%
  \BibitemOpen
  \bibfield  {author} {\bibinfo {author} {\bibfnamefont {S.~H.}\ \bibnamefont
  {Jacobsen}}, \bibinfo {author} {\bibfnamefont {J.~A.}\ \bibnamefont
  {Ouassou}}, \ and\ \bibinfo {author} {\bibfnamefont {J.}~\bibnamefont
  {Linder}},\ }\href {\doibase 10.1103/PhysRevB.92.024510} {\bibfield
  {journal} {\bibinfo  {journal} {Phys. Rev. B}\ }\textbf {\bibinfo {volume}
  {92}},\ \bibinfo {pages} {024510} (\bibinfo {year} {2015})}\BibitemShut
  {NoStop}%
\bibitem [{\citenamefont {Eschrig}(2000)}]{eschrig2000}%
  \BibitemOpen
  \bibfield  {author} {\bibinfo {author} {\bibfnamefont {M.}~\bibnamefont
  {Eschrig}},\ }\href {\doibase 10.1103/PhysRevB.61.9061} {\bibfield  {journal}
  {\bibinfo  {journal} {Phys. Rev. B}\ }\textbf {\bibinfo {volume} {61}},\
  \bibinfo {pages} {9061} (\bibinfo {year} {2000})}\BibitemShut {NoStop}%
\bibitem [{\citenamefont {Konstandin}\ \emph {et~al.}(2005)\citenamefont
  {Konstandin}, \citenamefont {Kopu},\ and\ \citenamefont
  {Eschrig}}]{konstandin2005}%
  \BibitemOpen
  \bibfield  {author} {\bibinfo {author} {\bibfnamefont {A.}~\bibnamefont
  {Konstandin}}, \bibinfo {author} {\bibfnamefont {J.}~\bibnamefont {Kopu}}, \
  and\ \bibinfo {author} {\bibfnamefont {M.}~\bibnamefont {Eschrig}},\ }\href
  {\doibase 10.1103/PhysRevB.72.140501} {\bibfield  {journal} {\bibinfo
  {journal} {Phys. Rev. B}\ }\textbf {\bibinfo {volume} {72}},\ \bibinfo
  {pages} {140501(R)} (\bibinfo {year} {2005})}\BibitemShut {NoStop}%
\bibitem [{\citenamefont {Ouassou}\ \emph {et~al.}(2016)\citenamefont
  {Ouassou}, \citenamefont {Bernardo}, \citenamefont {Robinson},\ and\
  \citenamefont {Linder}}]{ali2016}%
  \BibitemOpen
  \bibfield  {author} {\bibinfo {author} {\bibfnamefont {J.~A.}\ \bibnamefont
  {Ouassou}}, \bibinfo {author} {\bibfnamefont {A.~D.}\ \bibnamefont
  {Bernardo}}, \bibinfo {author} {\bibfnamefont {J.~W.~A.}\ \bibnamefont
  {Robinson}}, \ and\ \bibinfo {author} {\bibfnamefont {J.}~\bibnamefont
  {Linder}},\ }\href {\doibase 10.1038/srep29312} {\bibfield  {journal}
  {\bibinfo  {journal} {Sci. Rep.}\ }\textbf {\bibinfo {volume} {6}},\ \bibinfo
  {pages} {29312} (\bibinfo {year} {2016})}\BibitemShut {NoStop}%
\bibitem [{\citenamefont {Abrikosov}\ and\ \citenamefont
  {Gor'kov}(1960)}]{Abrikosov1960}%
  \BibitemOpen
  \bibfield  {author} {\bibinfo {author} {\bibfnamefont {A.~A.}\ \bibnamefont
  {Abrikosov}}\ and\ \bibinfo {author} {\bibfnamefont {L.~P.}\ \bibnamefont
  {Gor'kov}},\ }\href@noop {} {\bibfield  {journal} {\bibinfo  {journal} {Zhur.
  Eksptl'. i Teoret. Fiz.}\ }\textbf {\bibinfo {volume} {39}},\ \bibinfo
  {pages} {1781} (\bibinfo {year} {1960})}\BibitemShut {NoStop}%
\bibitem [{\citenamefont {Hauser}\ \emph {et~al.}(1966)\citenamefont {Hauser},
  \citenamefont {Theuerer},\ and\ \citenamefont {Werthamer}}]{hauser1966}%
  \BibitemOpen
  \bibfield  {author} {\bibinfo {author} {\bibfnamefont {J.~J.}\ \bibnamefont
  {Hauser}}, \bibinfo {author} {\bibfnamefont {H.~C.}\ \bibnamefont
  {Theuerer}}, \ and\ \bibinfo {author} {\bibfnamefont {N.~R.}\ \bibnamefont
  {Werthamer}},\ }\href {\doibase 10.1103/PhysRev.142.118} {\bibfield
  {journal} {\bibinfo  {journal} {Phys. Rev.}\ }\textbf {\bibinfo {volume}
  {142}},\ \bibinfo {pages} {118} (\bibinfo {year} {1966})}\BibitemShut
  {NoStop}%
\bibitem [{\citenamefont {Jarrell}(1988)}]{jarell1988}%
  \BibitemOpen
  \bibfield  {author} {\bibinfo {author} {\bibfnamefont {M.}~\bibnamefont
  {Jarrell}},\ }\href {\doibase 10.1103/PhysRevLett.61.2612} {\bibfield
  {journal} {\bibinfo  {journal} {Phys. Rev. Lett.}\ }\textbf {\bibinfo
  {volume} {61}},\ \bibinfo {pages} {2612} (\bibinfo {year}
  {1988})}\BibitemShut {NoStop}%
\bibitem [{\citenamefont {Hammer}\ \emph {et~al.}(2007)\citenamefont {Hammer},
  \citenamefont {Cuevas}, \citenamefont {Bergeret},\ and\ \citenamefont
  {Belzig}}]{hammer2007}%
  \BibitemOpen
  \bibfield  {author} {\bibinfo {author} {\bibfnamefont {J.~C.}\ \bibnamefont
  {Hammer}}, \bibinfo {author} {\bibfnamefont {J.~C.}\ \bibnamefont {Cuevas}},
  \bibinfo {author} {\bibfnamefont {F.~S.}\ \bibnamefont {Bergeret}}, \ and\
  \bibinfo {author} {\bibfnamefont {W.}~\bibnamefont {Belzig}},\ }\href
  {\doibase 10.1103/PhysRevB.76.064514} {\bibfield  {journal} {\bibinfo
  {journal} {Phys. Rev. B}\ }\textbf {\bibinfo {volume} {76}},\ \bibinfo
  {pages} {064514} (\bibinfo {year} {2007})}\BibitemShut {NoStop}%
\bibitem [{\citenamefont {Ouassou}\ \emph {et~al.}(2017)\citenamefont
  {Ouassou}, \citenamefont {Pal}, \citenamefont {Blamire}, \citenamefont
  {Eschrig},\ and\ \citenamefont {Linder}}]{ali2017}%
  \BibitemOpen
  \bibfield  {author} {\bibinfo {author} {\bibfnamefont {J.~A.}\ \bibnamefont
  {Ouassou}}, \bibinfo {author} {\bibfnamefont {A.}~\bibnamefont {Pal}},
  \bibinfo {author} {\bibfnamefont {M.}~\bibnamefont {Blamire}}, \bibinfo
  {author} {\bibfnamefont {M.}~\bibnamefont {Eschrig}}, \ and\ \bibinfo
  {author} {\bibfnamefont {J.}~\bibnamefont {Linder}},\ }\href {\doibase
  10.1038/s41598-017-01330-1} {\bibfield  {journal} {\bibinfo  {journal} {Sci.
  Rep.}\ }\textbf {\bibinfo {volume} {7}},\ \bibinfo {pages} {1932} (\bibinfo
  {year} {2017})}\BibitemShut {NoStop}%
\end{thebibliography}%

\end{document}